\documentstyle[12pt,aps]{revtex}
\begin{document}
\title
{Study of  $^3$He$(e,e')$ longitudinal response
functions
with the integral-transform method}
\author{V.Yu. Dobretsov and V.D. Efros}
\address{Russian Research Center "Kurchatov Institute", 123182 Moscow,
Russia}
\author{Bin Shao}
\address{Department of Physics, University of Pennsylvania,
Philadelphia, Pennsylvania 19104, USA}
\maketitle
\vspace{5mm}
\begin{abstract}
The method  of integral transforms is  first applied for  studying
the $^3$He  longitudinal
response functions. The
 transforms   are
calculated from localized bound-state-type solutions to an inhomogenous
Schr\"odinger-type three-body equation. The multipole expansion of the
solutions is considered.
  The whole
calculation is checked using a sum-rule test.
 Several versions of  local
$s$-wave spin-dependent potentials supplemented with a singlet $p$-wave
potential and with the proton-proton Coulomb interaction are used as a
two-nucleon input.
The conventional charge density operator involving free-nucleon
form factors
is utilized. The three-body equations are  solved with the help of  the
hyperspherical
expansion and a complete convergence is achieved.   The  final-state
interaction  is  thus fully taken into
account. The $q=300$ MeV/c and 500 MeV/c values are considered. It is found
that the contribution of the $T=3/2$ final states to
the problem is suppressed and it amounts about 15\%. This might be ascribed
to symmetry  requirements. The contributions of the $p$-wave $NN$ interaction
and of the Coulomb interaction are found to amount several per cent.
Uncertainty
due to different choices of $s$-wave $NN$ forces is of a similar magnitude
provided that the low-energy $NN$ data are properly described. The  results
are  compared  with  the  integral transforms
of the experimental response functions. For $q=300$ MeV/c experimental  and
theoretical results coincide
within their uncertainties. For $q=500$ MeV/c a noticeable difference is
detected.
\end{abstract}

Many calculations testify to the fact that the conventional form of
the  nuclear charge density is inapplicable for
 description of the
elastic
form factors of  three- and four-nucleon  nuclei at $q>2.5$  fm$^{-1}$
values.   However  the  elastic  scattering  occurs  with  a quite low
probability  at  such   $q$  values  and   some  non-typical
configurations  including  those  where  all  the  nucleons  are close
together may  contribute substantially.  In this connection  it
seems important to test a form of the nuclear 4-current in the
inelastic
processes and to study the ($e,e'$) response functions. This  requires
a proper account of the nuclear final-state interaction.

We study $(e,e')$ response functions
\begin{equation}
R(q,\omega)=\bar{\sum}_{M_0}\int df |<\psi_{f}|\hat
{O}|\psi_{0}>|^{2} \delta (E_{f}-E_{0}-\epsilon) \label{eq:defr}
\end{equation}
where $\epsilon$ is the nuclear excitation energy.
Direct evaluation of $R$
with a full account of the
final-state interaction
is quite complicated even
in the three-body case. Indeed,
it requires obtaining the whole set of the
final-state continuum wave functions $\psi_{f}$ at the same energy and
summing their contributions.
As far as we know
only one
 such an investigation was performed in the literature [1]. A local $s$-wave
central
potential [2] was utilized there while interaction in higher partial waves and
Coulomb interaction were disregarded.

In this paper we present  the first microscopical study of the $^3$He
longitudinal response functions $R_{l}$ with the help of the method of
integral transforms
[3-6]. While the final-state interaction is fully taken into account in our
approach we
avoid calculating the continuum-spectrum wave functions $\psi_{f}$. We explore
an
ability of the method in an A=3 problem. We study sensitivity of the results
to a choice of $NN$ force and we clarify a role of an interaction in higher
partial waves and that of the Coulomb interaction. We compare our results
with experiment [7,8] at $q=$ 300 and 500 MeV/c.

Define the reduced transition operator and the response function
\[  \tilde{O}=[\tilde{G}^{p}_{E}(Q)]^{-1}\hat{O},\,\,\,\,\,\,\,\,
\tilde{R}(q,\omega)=[\tilde{G}^{p}_{E}(Q)]^{-2}R(q,\omega). \]
Here $Q^2=q^2-\omega^2$ and [9]
$\tilde{G}^{p}_{E}(Q)=[1+Q^2/(4M^2)]^{-1/2}G^{p}_{E}(Q)$
where $G^{p}_{E}$ is the proton Sachs form factor.
We calculate the  integral transform of the response [3,10]
\begin{equation}
\Phi(q,\sigma)=\int_{\epsilon_{min}}^{\infty}(\sigma+\epsilon)^{-1}
\tilde{R}(q,\omega) d\epsilon   \label{eq:tran}
\end{equation}
instead of the response itself. We use the conventional
single-nucleon expression for the charge density $\hat{O}$. Then to a
very good approximation one can disregard the $\omega$-dependence of
the $\tilde{O}$ operator and use the expression
\begin{equation}
\tilde{O}({\bf q})=\sum_{j=1}^{A}[\frac{1-\tau_{zj}}{2}+
\frac{G^n_E(q)}
{G^p_E(q)}\frac{1+\tau_{zj}}{2}]e^{i{\bf q}{\bf r}_j'} \label {eq:op}
\end{equation}
where
$\mbox{\bf r}_j'={\bf r}_j-{\bf R}_{c.m}$. It has been
shown [3,10] that $\Phi(q,\sigma)$ can be calculated by first solving
for the {\em localized} solution to the following inhomogenous
equation
\begin{equation}
(H-E_{0}+\sigma)\tilde{\Psi}=\tilde{O}\psi_{0}  \label {eq:equ}
\end{equation}
where as in Eq. (1) $\psi_{0}$ is the ground-state wave function and
$E_{0}$
is the ground-state energy. In terms of $\tilde{\Psi}$
we have [3,10]
\begin{equation}
\Phi(q,\sigma)=<\tilde{\Psi}|\tilde{O}\psi_{0}>-\sigma^{-1}
\tilde{R}_{el}   \label {eq:phi}
\end{equation}
where $\tilde{R}_{el}$ is the elastic contribution to the response.

The solution to Eq. (\ref {eq:equ}) is much easier to obtain
than the functions $\psi_{f}$ entering Eq. (\ref{eq:defr}).  Indeed, in
contrast
to the latter functions there is no need to impose the complicated
large-distance boundary conditions in order to fix a solution. The only
condition of vanishing of the solution at large distances suffices.
Therefore methods that are used in
solving bound-state problems can be utilized here. Below we
use real $\sigma$ values in Eqs. (\ref{eq:tran}), (\ref{eq:equ}).

We have two possible ways to connect our theoretical calculations
with experiment. One way [10] is to
compare $\Phi(q,\sigma)$ with the same quantity
obtained from the experimental $\tilde{R}(q,\epsilon)$ using Eq. (2).
Another way [3] is to consider Eq. (\ref{eq:tran}) as the integral equation
and
invert to obtain theoretical $\tilde{R}(q,\omega)$ and
then compare the responses themselves.  In the
present work we use the first approach. This provides one only with a limited
information since some $R$ features are smeared out at performing the
transform (\ref{eq:tran}). Nevertheless, this still allows testing
the $\hat{O}$ transition
operator in Eq. (\ref{eq:defr}). If, for example, the nucleon form factors
should be
modified inside the nucleus this would produce a change in $\Phi(q,\sigma)$
of a magnitude comparable with the change in $(G_E^p)^2$.

Concerning the
inversion problem, another version of the method [6] better suits for this
purpose. The many-body equation to be solved in this version is similar
to Eq. (\ref{eq:equ}) yet. The only difference is that complex $\sigma$ values
are utilized. It is shown below that the left-hand side of Eq. (\ref{eq:tran})
can be obtained from  Eq. (\ref{eq:equ}) with high accuracy and we achieved
comparable accuracy for the complex $\sigma$ values as well.

The calculations are performed at $q=300$ and 500 MeV/c and they are
compared with the Saclay [7] and Bates
[8] experimental results.
In this first calculation we use four versions of effective central local
$s$-wave spin-dependent $NN$ potentials [2,11,12] supplemented with
a realistic singlet $p$-wave $NN$ potential [13] and with the proton-proton
Coulomb interaction, Only the components of latter interaction which are
diagonal in the isospin $T=$1/2, 3/2 quantum numbers are retained in the
calculation. Even these components prove to contribute little to the results,
see below.

Under these assumptions on the nuclear dynamics, Eq. (\ref {eq:equ})
is split
into independent sets of equations with a given orbital momentum $L$,
and isospin $T$ of the system.  It is convenient to calculate the
right-hand sides of these equations in the following way.  Since
$\psi_0$ has $L=0$ then only the components $\sim
\sum_{M_L}Y_{LM_L}(\hat{{\bf r}}_j')Y_{LM_L}^*(\hat{{\bf q}})$
from the expansion of
$\exp(i{\bf q}{\bf r}_j')$ from Eq. (\ref{eq:op})
contribute to the problem for a given $L$ value. $R$ is independent of a
${\bf q}$ direction
due to averaging over $M_0$ in Eq. (1). Let ${\bf
q}$ be directed along the $z$ axis. Then only the components
with $M_L=0$ give non-zero contributions and hence only the
components of $\tilde{\Psi}$ with $M_L=0$ are different from zero. We
have
\begin{eqnarray}
(H-E_{0}+\sigma)\tilde{\Psi}_{LT}=\tilde{O}_{LT}\psi_{0},
\label {eq:eqlt}  \\
\tilde{O}_{LT}=\sum_{M_T}|TM_T><TM_T
|\sum_{j=1}^3[\frac{1-\tau_{zj}}{2}
+\frac{G^n_E(q)}
{G^p_E(q)}\frac{1+\tau_{zj}}{2}]j_L(qr_j)Y_{L0}(\hat{{\bf
 r}}_j'),   \label {eq:oplt}\\
\Phi_{LT}(q,\sigma)=<\tilde{\Psi}_{LT}|\tilde{O}_{
LT}\psi_0>-\delta_{L0}\sigma^{-1}R_{el},  \\
\Phi(q,\sigma)=4\pi\sum_{L=0}^{\infty}(2L+1)\sum_{T=1/2,\,3/2}\Phi_{LT}
(q,\sigma). \label {eq:suml}
\end{eqnarray}
The functions $\tilde{\Psi}_{LT}$ have the same spin $S=1/2$
as $\psi_0$.

We solve Eqs. (\ref {eq:eqlt})  using the  hyperspherical
expansion. Eqs. (6) turn into  sets of algebraic equations.
Denote $K$ the hyperspherical momentum and $[f]$
the type
of symmetry of the spatial components of the basis functions.
Our basis functions are of the form
\begin{equation}
R_n(\rho)\Gamma_{KLM_LSM_STM_{T}
;i}^{[f]\mu_f}(\Omega,\sigma_{zj},
\tau_{zj}).   \label {eq:k}
\end{equation}
Here $\rho$ is the hyperradial variable and $\Omega$ are hyperangular
variables. The index $i$ enumerates the functions with the same other quantum
numbers. We use the Laguerre-type hyperradial basis functions [14]. We have
developed  a  computer  code  to   construct  complete  sets  of   the
$\Gamma$ basis functions from Eq. (10)  with arbitrary  quantum numbers using
the
the Raynal-Revai   transformation   [15].   The   coefficients   of   this
transformation  are  evaluated  using  the  recurrent  formula  of the
$K\rightarrow K+2$ type  [16].

The $L$ values
up to $L_{max}=8$ are retained which provides convergence in Eq. (9).
The net number of the basis functions  with  the  same  $K,L,S,T$  and $[f]$
values grows linearly
with $K$ which in general leads to  systems of linear equations of
high dimensions. But in fact these dimensions can be considerably
reduced.
For this purpose we use the fact
that nuclear forces we utilize here  only act in the two $NN$
orbital  states. Namely,
we specify orthonormalized basis functions in such a way [17] that
only for two
functions from each set of functions with the same $K,L,S,T$ and $[f]$
the matrix elements of $NN$ force are different from zero.
Only such functions are coupled together in Eqs. (6). Other functions
do not include $s$ or $p$ components for each nucleon pair. They lead to
decoupled equations in Eqs. (6) and they correspond to free-motion final
states.

We retained the basis functions (10) with
$K\leq K_{max}=30$ in our calculations. This provides practical convergence of
the results. The convergence trends are illustrated in
Fig. 1 where dependence on $K_{max}$ of the calculated
$\Phi_{L=2,T=1/2}(q,\sigma)$ contribution
is shown  at $q=300$ MeV/c and $\sigma=1$ MeV as an example.
In order to check stability of our results with respect to $K_{max}$ we
performed their
extrapolations
to $K_{max}=\infty$. This was done using three-parameter
fits of the form
$\Phi(K_{max})=\Phi(\infty)-c/K_{max}^{\gamma}$. Extrapolated values
$\Phi(\infty)$  differ from $\Phi(K_{max}=30)$ less than by 1\%.
Convergence with respect to hyperradial
basis functions from Eq. (\ref {eq:k}) is achieved, too. The maximum number of
coupled
basis states equaled 861 at solving  Eqs. (6). The calculated transforms are
smooth functions similar to shown in Fig. 1a from Ref. 4 for the deutron case.

We make a comment concerning calculation of $\Phi$ at small $\sigma$
values.  It is necessary to avoid large cancellations in the
right-hand side of Eq. (\ref {eq:phi}).  This is achieved if one uses
in Eq. (\ref {eq:eqlt}) for $L=0$, $T=1/2$  the same $K_{max}$ value as
that at calculating $\psi_0$. Then the pole terms cancel exactly.

There exists a test which enables us to check the calculation as a
whole. Namely,  the leading term  of $\Phi(q,\sigma)$ at high
$\sigma$ values behaves as $\sigma^{-1}$. This term can be calculated
independently using the sum rule,
\begin{eqnarray}
\lim_{\sigma \rightarrow \infty}\sigma \Phi(q,\sigma)=
\int_{\epsilon_{min}}^{\infty}\tilde{R}(q,\omega) d\epsilon
=\bar {\sum}_{M_0}<\psi_0|\tilde{O}^{\dag}
\tilde{O}|\psi_0>-\tilde{R}_{el}.  \label {eq:ps}
\end{eqnarray}
This allows one to check the right-hand side of Eq. (\ref {eq:suml}).
Besides checking the calculation the test allows one to
verify whether at high $\sigma$ values the results  are stable
against increasing $K_{max}$ and $L_{max}$.\footnote {This test, however, is
not applicable for checking the matrix elements of the Hamiltonian in
Eqs. (6).
(When $\sigma$ tends to infinity these matrix elements can be neglected.)
In a preliminary version of this work [18]
a considerable disagreement with experiment was found. This was due to
an
algebraic error. For mixed symmetry states the matrix elements of kinetic
energy were erroneously twice as large as correct ones.}

Various contributions to Eq. (9) are just the integral transforms of the
corresponding contributions to $R$. We note that the $T=3/2$ contributions
prove to be suppressed in comparison with the $T=1/2$ ones. The net relative
$T=3/2$ contribution to $\Phi(q,\sigma)$ ranges between 12 and 17\% for all
the $\sigma$ values and the two $q$ values considered. The contributions to
the initial responses $R$ should be of a similar magnitude. The reason for
the suppression may
be in that the spatial component of
the final-state wave function which are symmetrical with respect to particle
interchanges are present at $T=1/2$ only. These components may provide more
internucleon attraction than those of the other symmetries which
increases amplitudes of the final-state wave functions inside the
reaction zone.

We estimated an influence on the results of $NN$ interactions in higher
partial waves and of the proton Coulomb interaction. The results are shown in
Fig. 2 for $q=300$ MeV/c. We performed the calculation for the MT(I+III)
potential supplemented with the realistic $p$-wave singlet interaction from
Ref. 13 and with the Coulomb interaction. Then we switched off the
$p$-wave interaction. Curve 1 shows the relative change in the results. The
contribution from the triplet $p$-wave force is believed to be of the same
size. After that we switched off the Coulomb interaction in addition. This
produces Curve 2. It is seen that the contributions of the $p$-wave
interaction and of the Coulomb interaction do not exceed several per cent or
standard  uncertainties in $\Phi$ functions obtained from experimental data.
For $q=500$ MeV/c these contributions are even much smaller.

Furthermore, we studied dependence of the results on the choice of the $NN$
force.
Central local s-wave $NN$ forces
utilized in our calculation include
the MT(I+III) [2], S2,S3 [11] and EH [12] potentials. The MT(I+III) and S2
potentials reproduce
the $NN$ low-energy properties and $s$-wave $NN$ phases up to high
energies.  The S3 potential fits the low-energy data and yields nearly
correct values for the binding energies and rms radii of $^3$He and $^4$He.
The EH potential fits the $s$-wave $NN$
phases up to high
energies  but it does not reproduce properly the low-energy
$NN$ data (see Ref. 11). As above   all these forces
were supplemented with the realistic $p$-wave singlet interaction [13] and
with the proton Coulomb interaction in  our calculations. Curve 3 in
Fig. 2 represents the relative difference in $\Phi$ between S2 and MT(I+III)
potentials. In case of S3 and MT(I+III) potentials such a difference is
approximately of the same value and of opposite sign.  These differences do
not exceed several per cent. Curve 4 represents such a difference between EH
and  MT(I+III)
potentials. The latter difference exceeds ten per cent at small $\sigma$
values which
corresponds to substantial differences in low-energy parts of the responses.
This
can be attributed to the above-mentioned drawback of the EH potential.
For $q=500$ MeV/c  differences in the results for different potentials
become considerably smaller.

Before comparing our calculation with experiment
we make some comments on obtaining the integral transforms,
Eq. (\ref{eq:tran})),
of the experimental responses.
In order to perform the
integration in Eq. (2) with a sufficient accuracy and in particular
to estimate the contribution from the unavailable high-$\epsilon$
tails of the spectra we approximate the experimental $\tilde{R}$
functions by the following
analytical expressions: $a(\omega-\omega_{thresh})^{1/2}$ in the low
$\omega$ region $\omega_{thresh}\leq\omega\leq\omega_1$, then
$\sum_{n=0}^{N_{b}}b_n\omega^n$ in the region of the peak
$\omega_1\leq\omega\leq\omega_2$ and
$\sum_{n=0}^{N_{c}}c_n(\omega_2/\omega)^{\alpha+n}$ in the region
beyond the peak. The parameters $a,b_n,c_n,\alpha$ and $\omega_2$ are
chosen from the least-square procedure with  additional requirements imposed
of
continuity of the fitting spectra and their first derivatives at
$\omega_1$ and $\omega_2$ points. It turns out that a
good description at  quite wide ranges
$\omega_2\leq\omega\leq\omega_{max}$ of the spectra beyond the peaks
are provided with a single term [19] $\sim \omega^{-\alpha} $. (In case,
say, exponentially decreasing tail-terms the description is worse.)
The best $\alpha$ values range between about 4 and 5 for all the $q$
values considered. Similar $\alpha$ values were found [20] in the
$^4$He case. We extrapolate the fitted spectra  beyond the
$\omega_{max}$ values in order to take into account the contributions
from the unavailable tails. In case of data from Ref. [7] these contributions
proved
to be quite small. They reach their maxima at high $\sigma$
values where they are between 1 and 2\%. In case of q=500 MeV data from
Ref. [8]
we did not succeed in producing  stable extrapolations.

In Fig. 3
we compare  the calculated $\Phi$ values and the values deduced from
experiment. The theoretical calculation was done MT(I+III) $NN$ potential [2]
supplemented as above with
$p$-wave singlet $NN$ interaction and with the proton
Coulomb interaction. The relative differences between the theoretical
and experimental $\Phi$ values are shown.
Taking into account 5\% systematic uncertainties of the experimental
data and also above-considered
uncertainties of theoretical calculations  one can say that
that there is no significant difference between experiment and theory at
$q=300$ MeV/c. The deviation at $q=$500 MeV/c may be considered as
significant. Detectable differences betweem experiment and theory at such
$q$ values
for another set [8] of data were also obtained in Ref. [1] where the responses
were calculated directly.
They  may be attributed to relativistic effects.

In conclusion, we applied the method of integral transforms for studying
$^3$He longitudinal response functions. This requires solving for a
localized solution to an inhomogenous  Schr\"odinger-type three-body
equation. We elaborated techniques for this purpose and we obtained  accurate
solutions using central local $NN$ potentials and the conventional expression
for the nuclear charge density. We found that $NN$ interaction in higher
patial waves and the proton Coulomb interaction play a minor role in the
problem. Uncertainties in the $s$-wave
$NN$ force  proved to be not substantial as well provided that the low-energy
$NN$-data are properly described. We compared the calculated integral
transforms with experiment and we found that
at $q=$300 MeV/c they agree with each other within
their uncertainties. For $q=$500 MeV/c noticeable deviations are found. The
results obtained make it possible to solve the corresponding problem in the
quite important $\alpha$-particle case, in particular. In addition, we found
that
the final-state $T=3/2$ contributions to the problem are suppressed.

We are grateful to K. Dow and C. Marchand for providing us with the
experimental data. Helpful discussions with G. Do Dang and
Yu. E. Pokrovsky are gratefully acknowledged. The work was supported
by the Russian Fund for Fundamental Research under Grant No.
93-02-14405.
\pagebreak
\\
{\bf References}\\
$[1]$ E. van Meijgaard and J. A. Tjon, Phys. Rev. C {\bf 45}, 1463 (1992);
Phys Lett {\bf B228}, 307 (1989)
\\
$[2]$ R. A. Malfliet and J. A. Tjon, Nucl. Phys. {\bf A127}, 161
(1969).\\
$[3]$ V. D. \'Efros, Yad. Fiz. {\bf 41}, 1498 (1985) [Sov. J. Nucl.
Phys. {\bf 41}, 949 (1985)].\\
$[4]$ V. D. Efros, W. Leidemann and G. Orlandini, Few-Body Syst.,
{\bf 14}, 151 (1993).\\
$[5]$ V. D. \'Efros, Yad. Fiz. {\bf 56}, No. 7, 22 (1993) [Phys. At. Nucl.
{\bf 56}, 869 (1993)] ({\em Proc. Int. Woorkshop})\\
$[6]$  V. D. Efros, W. Leidemann and G. Orlandini,
Phys. Lett. B., in press. \\
$[7]$ C. Marchand et al., Phys. Lett. {\bf 153B}, 29 (1985) and
private communication.\\
$[8]$ K. Dow et al. Phys. Rev. Lett. {\bf 61}, 1706 (1988); K. Dow,
Ph.D. thesis, MIT, 1987.\\
$[9]$ J. L. Friar, Phys. Lett. {\bf 43B}, 108 (1973); D. R. Yennie,
M. Levy and D. G. Ravenhall, Rev. Mod. Phys. {\bf 29}, 144 (1957).\\
$[10]$ V. D. \'Efros, Ukr. Fiz. Zh. {\bf 25}, 907 (1980)
[Ukr. Phys. J.] (sect. 2C).\\
$[11]$ I. R. Afnan and Y. C. Tang, Phys. Rev. {\bf 175}, 1337 (1968).\\
$[12]$ H. Eikemeier and H. H. Hackenbroich, Z. Phys. {\bf 195}, 412
(1966).\\
$[13]$ R. de Tourreil, B. Rouben and D. W. L. Sprung, Nucl. Phys.
{\bf A242}, 445 (1973).\\
$[14]$ G. Erens, J.L. Visschers and R. van Wageningen, Ann. Phys. {\bf 67},
461 (1971).\\
$[15]$ J. Raynal and J. Revai, Nuovo Cim. {\bf A68}, 612 (1970).\\
$[16]$ Ya. A. Smorodinskii and V. D. \'Efros, Yad. Fiz. {\bf 17},
210 (1973) [Sov. J. Nucl. Phys. {\bf 17}, 107 (1973)]. \\
$[17]$ V. D. \'Efros, Yad. Fiz. {\bf 15}, 226 (1972) [Sov. J. Nucl.
Phys. {\bf 15}, 128 (1972)].\\
$[18]$ V. Yu. Dobretsov, V. D. Efros and Bin Shao, Kurchatov preprint
IAE-5650/2, 1993.\\
$[19]$ G. Orlandini, M. Traini, Phys. Rev. {\bf C31}, 280 (1985).\\
$[20]$ A. Yu. Buki et al., Kurchatov preprint IAE-5397/2, 1991;
V. D. Efros, Few-Body Syst. Suppl. 6, 506 (1992).\\
\pagebreak
\\
\centerline{CAPTION TO FIGURES }

\vspace{10mm}
\noindent
FIG. 1. Dependence of $\Phi_{LT}(q,\sigma)$ on $K_{max}$ at $L=2$, $T=1/3$,
$q=300$ MeV/c and $\sigma=1$ MeV.

\vspace{10mm}

\noindent
FIG. 2. Relative differences in $\Phi(q,\sigma)$ between MT(I+III)
potential [2] supplemented with  $p$-wave singlet interaction plus Coulomb
proton interaction
and other $NN$ inputs. Curve 1 - $p$-wave interaction switched off.
Curve 2 -
Coulomb interaction switched off in addition. Curve 3 - a difference with
the S2 $NN$ force [11].
Curve 4 - a difference with EH $NN$ force [12]. (In the last two cases
$p$-wave singlet
interaction
and Coulomb interaction are included in the calculation.)

\vspace{10mm}

\noindent
Fig.3. Comparison of the calculated transforms $\Phi_{th}(q,\sigma)$ with
experiment. Curve 1 is for experimental data from Ref. [7] for $q=300$ MeV/c.
Curve 2 is for experimental data from Ref. [8] for $q=300$ MeV/c. Curve 3 is
for experimental data from Ref. [8] for $q=500$ MeV/c.
\end{document}